\newif\ifproblem
\newif\ifobservation
\newif\iftimesok

\makeatletter
\def\IfStandaloneCheck{\def\next{aipcheck}
  \edef\currjob{\jobname}
  \edef\next{\meaning\next}
  \edef\currjob{\meaning\currjob}
  \ifx\currjob\next
    \expandafter\@firstoftwo
  \else
    \expandafter\@secondoftwo
  \fi
}
\makeatother

\ifx\documentclass\undefined
 \expandafter\stop
\else
\fi

\def\next#1/#2/#3\next{#1#2}
\ifnum\expandafter\next\fmtversion\next<199612 \relax
 \expandafter\stop
\else
 \ifnum\expandafter\next\fmtversion\next<199806 \relax
   \typein{* Type <return> to continue ...}
   \problemtrue
 \else
 \fi
\fi

\IfFileExists{aipproc.cls}
    {
     \typeout{* ... ok }
    }
    {
     \typeout{* ... not found! }
     \typeout{*}
     \typeout{* Sorry this is a fatal error:}
     \typeout{*}
     \typeout{* Before you can use the aipproc class you have to unpack}
     \typeout{* it from the documented source.}
     \typeout{*}
     \typeout{* Run LaTeX on the file 'aipproc.ins', e.g.,}
     \typeout{*}
     \typeout{* \space\space latex aipproc.ins}
     \typeout{*}
     \typeout{* or whatever is necessary on your installation to process}
     \typeout{* a file with LaTeX. This should unpack a number of files for you:}
     \typeout{*}
     \typeout{* aipproc.cls \space and \space aip-*.clo}
     \typeout{*}
     \typeout{* After that retry processing this guide.}
     \typeout{*}
     \stop
}

\IfFileExists{aipxfm.sty}
    {
     \typeout{* ... ok }
    }
    {
     \typeout{* ... not found! }
     \typeout{*}
     \typeout{* Sorry this is a fatal error:}
     \typeout{*}
     \typeout{* The aipxfm.sty file which is part of the aipproc distribution}
     \typeout{* must be installed in a directory which is searched by LaTeX.}
     \typeout{*}
     \typeout{* Please install this file and retry.}
     \typeout{*}
     \stop
}

\IfFileExists{aip-8s.clo}
    {
     \typeout{* ... ok }
    }
    {
     \typeout{* ... not found! }
     \typeout{*}
     \typeout{* Sorry this is a fatal error:}
     \typeout{*}
     \typeout{* The aip-8s.clo file which is part of the aipproc distribution}
     \typeout{* must be installed in a directory which is searched by LaTeX.}
     \typeout{*}
     \typeout{* Please install this file and retry.}
     \typeout{*}
     \stop
}

\IfFileExists{aip-8d.clo}
    {
     \typeout{* ... ok }
    }
    {
     \typeout{* ... not found! }
     \typeout{*}
     \typeout{* Sorry this is a fatal error:}
     \typeout{*}
     \typeout{* The aip-8d.clo file which is part of the aipproc distribution}
     \typeout{* must be installed in a directory which is searched by LaTeX.}
     \typeout{*}
     \typeout{* Please install this file and retry.}
     \typeout{*}
     \stop
}

\IfFileExists{aip-6s.clo}
    {
     \typeout{* ... ok }
    }
    {
     \typeout{* ... not found! }
     \typeout{*}
     \typeout{* Sorry this is a fatal error:}
     \typeout{*}
     \typeout{* The aip-6s.clo file which is part of the aipproc distribution}
     \typeout{* must be installed in a directory which is searched by LaTeX.}
     \typeout{*}
     \typeout{* Please install this file and retry.}
     \typeout{*}
     \stop
}

\IfFileExists{fixltx2e.sty}
    {
     \typeout{* ... ok }
    }
    {
     \typeout{* ... not found, trying fix2col.sty instead ... }
     \typeout{*}
     \IfFileExists{fix2col.sty}
         {
          \typeout{* ... ok }
         }
         {
          \typeout{* ... not found! }
          \typeout{*}
          \typeout{* Sorry this is a fatal error:}
          \typeout{*}
          \typeout{* Your LaTeX distribution contains neither fixltx2e.sty}
          \typeout{* nor fix2col.sty.}
          \typeout{*}
          \typeout{* This means that it is either too old or incompletely}
          \typeout{* installed.}
          \typeout{*}
          \typeout{* fixltx2e.sty is part of the standard LaTeX distribution}
          \typeout{* since 1999; fix2col.sty is an earlier version of this}
          \typeout{* package.}
          \typeout{*}
          \typeout{* Best solution is to get the latest LaTeX distribution.}
          \typeout{* If this is impossible for you, download fix2col.sty.}
          \typeout{* You can get this software from a CTAN host.}
          \typeout{* Refer to http://www.ctan.org and search for "fix2col".}
          \typeout{*}
          \typeout{* After you have updated your LaTeX distribution}
          \typeout{* retry processing this guide.}
          \stop
     }
}

\IfFileExists{fontenc.sty}
    {
     \typeout{* ... ok }
    }
    {
     \typeout{* ... not found! }
     \typeout{*}
     \typeout{* Sorry this is a fatal error:}
     \typeout{*}
     \typeout{* The fontenc package, which is part of standard LaTeX}
     \typeout{* (base distribution) has to be installed at the site to}
     \typeout{* run the aipproc class.}
     \typeout{*}
     \typeout{* The fact that it cannot be found either means that}
     \typeout{* this LaTeX release is too old or that it was installed}
     \typeout{* improperly.}
     \typeout{*}
     \typeout{* Please make sure that your version of LaTeX is okay}
     \typeout{* before attempting to use this class. The LaTeX distribution}
     \typeout{* contains the file "ltxcheck.tex" which can be used to}
     \typeout{* test the basic functionality and integrity of your installation.}
     \typeout{*}
     \stop
    }

\IfFileExists{calc.sty}
    {
     \typeout{* ... ok }
    }
    {
     \typeout{* ... not found! }
     \typeout{*}
     \typeout{* Sorry this is a fatal error:}
     \typeout{*}
     \typeout{* The calc package, which is part of standard LaTeX}
     \typeout{* (tool distribution) has to be installed at the site}
     \typeout{* to run the aipproc class.}
     \typeout{*}
     \typeout{* The fact that it cannot be found either means that}
     \typeout{* this LaTeX release is too old or that it was installed}
     \typeout{* only in parts.}
     \typeout{*}
     \typeout{* Please make sure that the tools distribution of LaTeX}
     \typeout{* is installed before attempting to use this class.}
     \typeout{*}
     \typeout{* (You might be able to get calc.sty separately for your}
     \typeout{* installation if you are unable to upgrade to a recent}
     \typeout{* distribution for some reason.)}
     \typeout{*}
     \stop
    }

\IfFileExists{varioref.sty}
    {
     \typeout{* ... ok }
     
    }
    {
     \typeout{* ... not found! }
     \typeout{*}
     \typeout{* Problem detected:}
     \typeout{*}
     \typeout{* The varioref package, which is part of standard LaTeX}
     \typeout{* (tool distribution) is not installed at this site.}
     \typeout{*}
     \typeout{* The fact that it cannot be found either means that}
     \typeout{* this LaTeX release is too old or that it was installed}
     \typeout{* only in parts.}
     \typeout{*}
     \typeout{* You can use the aipproc class without this package but }
     \typeout{* you cannot make use of the options "varioref" or "nonvarioref".}
     \typeout{*}
     \typeout{* Please also note that the aipguide.tex documentation}
     \typeout{* normally uses the "varioref" option to show its}
     \typeout{* effects (which  will now fail).}
     \typeout{*}
     \typein{* Type <return> to continue ...}
     \problemtrue
     
     \let\vpageref\pageref
     
    }

\IfFileExists{times.sty}
    {
     \begingroup
       \RequirePackage{times}
       \global\expandafter\let\csname ver@times.sty\endcsname\relax    
       \long\def\next{ptm}
       \ifx\rmdefault\next
         \typeout{* ... ok }
         
         \endgroup
         \timesoktrue
       \else
         \endgroup
     \typeout{* ... obsolete! }
     \typeout{*}
     \typeout{* Serious problem detected:}
     \typeout{*}
     \typeout{* The times package, which is part of standard LaTeX}
     \typeout{* (psnfss distribution) is obsolete at this site.}
     \typeout{*}
     \typeout{* The fact that it contains incorrect code either means that}
     \typeout{* this LaTeX release is too old or that it was installed}
     \typeout{* only in parts with old files remaining!}
     \typeout{*}
     \typeout{* You can use the aipproc class without this package but}
     \typeout{* you have to specify the option "cmfonts" which result in}
     \typeout{* documents which are not conforming to the AIP layout specification!}
     \typeout{*}
     \typeout{* You can also try using the class in the following way:}
     \typeout{*}
     \typeout{* \space\space \string\documentclass[cmfonts]{aipproc}}
     \typeout{* \space\space \string\usepackage{times}}
     \typeout{* \space\space ...}
     \typeout{*}
     \typeout{* With luck this will result in Times Roman output but chances}
     \typeout{* are that you will get a larger number of error messages in}
     \typeout{* which case you have to remove the \string\usepackage declaration.}
     \typeout{*}
     \typein{* Type <return> to continue ...}
          \problemtrue
          
       \fi
    }
    {
     \typeout{* ... not found! }
     \typeout{*}
     \typeout{* Serious problem detected:}
     \typeout{*}
     \typeout{* The times package, which is part of standard LaTeX}
     \typeout{* (psnfss distribution) can not be found.}
     \typeout{*}
     \typeout{* The fact that this package cannot be found either means that}
     \typeout{* this LaTeX release is too old or that it was installed}
     \typeout{* only in parts!}
     \typeout{*}
     \typeout{* You can use the aipproc class without this package but }
     \typeout{* you have to specify the option "cmfonts" which result in}
     \typeout{* documents which are not conforming to the AIP layout specification!}
     \typeout{*}
     \typein{* Type <return> to continue ...}
     \problemtrue
     
    }

\iftimesok 

\IfFileExists{t1ptm.fd}
    {
     \typeout{* ... ok }
    }
    {
     \typeout{* ... not found, trying T1ptm.fd ... }
     \IfFileExists{T1ptm.fd}
          {
           \typeout{* ... ok }
          }
          {
           \typeout{* ... not found}
           \typeout{* Serious problem detected:}
           \typeout{*}
           \typeout{* The times package, which is part of standard LaTeX}
           \typeout{* (psnfss distribution) is available but the corresponding}
           \typeout{* .fd file (defining how to load Times Roman) is missing.}
           \typeout{*}
           \typeout{* The fact that this package is only partially installed}
           \typeout{* means that you LaTeX installation is unable to use Times}
           \typeout{* Roman fonts!}
           \typeout{*}
           \typeout{* You can use the aipproc class without this package but }
           \typeout{* you have to specify the option "cmfonts" which result in}
           \typeout{* documents which are not conforming to the AIP layout}
           \typeout{* specification!}
           \typeout{*}
           \typein{* Type <return> to continue ...}
           \problemtrue
           \timesokfalse
           
          }
    }

\fi

\newcommand\CheckFDFile[3]{%
  \typeout{*}
  \typeout{* Looking for #1#3.fd or #2#3.fd ... }
  \IfFileExists{#1#3.fd}
    {
     \typeout{* ... ok }
    }
    {
     \IfFileExists{#2#3.fd}
      {
       \typeout{* ... ok }
      }
      {\problemtrue
       \typeout{* ... not found! }
      }
    }
}

\iftimesok 

\IfFileExists{mathptm.sty}
    {
     \typeout{* ... ok }
     \CheckFDFile{ot1}{OT1}{ptmcm}
     \CheckFDFile{oml}{OML}{ptmcm}
     \CheckFDFile{oms}{OMS}{pzccm}
     \CheckFDFile{omx}{OMX}{psycm}
     \ifproblem
      \typeout{*}
      \typeout{* Problem detected:}
      \typeout{*}
      \typeout{* The mathptm package, which is part of standard LaTeX}
      \typeout{* (psnfss distribution) was found but some or all of its}
      \typeout{* support files describing which fonts to load are missing!}
      \typeout{*}
      \typeout{*}
      \typeout{* The fact that this package is only partially installed}
      \typeout{* means that the mathptm package cannot be used!}
      \typeout{*}
      \typeout{* You can use the aipproc class without this package but }
      \typeout{* you have to specify the option "nomathfonts" so that}
      \typeout{* math formulas will be typeset using Computer Modern.}
      \typeout{*}
      \typein{* Type <return> to continue ...}
      \problemtrue
      
     \else
      \typeout{*}
      \typeout{* Looking for mathptmx.sty ... }
      \IfFileExists{mathptmx.sty}
       {
        \typeout{* ... ok }
        \CheckFDFile{ot1}{OT1}{ztmcm}
        \CheckFDFile{oml}{OML}{ztmcm}
        \CheckFDFile{oms}{OMS}{ztmcm}
        \CheckFDFile{omx}{OMX}{ztmcm}
        \ifproblem
          \typeout{*}
          \typeout{* Problem detected:}
          \typeout{*}
          \typeout{* The mathptmx package, which is part of standard LaTeX}
          \typeout{* (psnfss distribution) was found but some or all of its}
          \typeout{* support files describing which fonts to load are missing!}
          \typeout{*}
          \typeout{*}
          \typeout{* The fact that this package is only partially installed}
          \typeout{* means that the mathptmx package cannot be used!}
          \typeout{*}
          \typeout{* You can use the aipproc class without this package but }
          \typeout{* you have to specify the option "mathptm" (no x) so that}
          \typeout{* math formulas use the older version with upright greek letters.}
          \typeout{*}
          \typein{* Type <return> to continue ...}
          \problemtrue
          
        \fi
       }
       {
        \typeout{* ... not found! }
        \typeout{*}
        \typeout{* Problem detected:}
        \typeout{*}
        \typeout{* The mathptmx package, which is part of standard LaTeX}
        \typeout{* (psnfss distribution) can not be found.}
        \typeout{*}
        \typeout{* This is unfortunate but not a disaster as the older}
        \typeout{* version of the package "mathptm" (no x) seems to exist.}
        \typeout{*}
        \typeout{* You can use the aipproc class without this package but }
        \typeout{* you have to specify the option "mathptm" so that}
        \typeout{* math formulas use the older version with upright greek letters.}
        \typeout{*}
        \typein{* Type <return> to continue ...}
        \problemtrue
        
       }
      \fi
    }
    {
     \typeout{* ... not found! }
     \typeout{*}
     \typeout{* Problem detected:}
     \typeout{*}
     \typeout{* The mathptm package, which is part of standard LaTeX}
     \typeout{* (psnfss distribution) can not be found.}
     \typeout{*}
     \typeout{* The fact that this package cannot be found either means that}
     \typeout{* this LaTeX release is too old or that it was installed}
     \typeout{* only in parts!}
     \typeout{*}
     \typeout{* You can use the aipproc class without this package but }
     \typeout{* you have to specify the option "nomathfonts" so that}
     \typeout{* math formulas will be typeset using Computer Modern.}
     \typeout{*}
     \typein{* Type <return> to continue ...}
     \problemtrue
     
    }

\IfFileExists{mathtime.sty}
    {
     \typeout{* ... ok }
    }
    {
     \typeout{* ... not found! }
     \typeout{*}
     \typeout{* The mathime package can not be found.}
     \typeout{*}
     \typeout{* This is not a real problem but an observation,}
     \typeout{* because this package is only of interest}
     \typeout{* if you own the commerical MathTime fonts.}
     \typeout{*}
     \typeout{* You can use the aipproc class without this package but }
     \typeout{* you cannot use the "mathtime" option of the class.}
     \typeout{*}
     \observationtrue
    }
\IfFileExists{mtpro.sty}
    {
     \typeout{* ... ok }
    }
    {
     \typeout{* ... not found! }
     \typeout{*}
     \typeout{* The mtpro package can not be found.}
     \typeout{*}
     \typeout{* This is not a real problem but an observation,}
     \typeout{* because this package is only of interest}
     \typeout{* if you own the commerical MathTime Professional fonts.}
     \typeout{*}
     \typeout{* You can use the aipproc class without this package but }
     \typeout{* you cannot use the "mtpro" option of the class.}
     \typeout{*}
     \observationtrue
    }
\else
\fi 

\IfFileExists{graphicx.sty}
    {
     \typeout{* ... ok }
    }
    {
     \typeout{* ... not found! }
     \typeout{*}
     \typeout{* Problem detected:}
     \typeout{*}
     \typeout{* The graphics package, which is part of standard LaTeX}
     \typeout{* (graphics distribution) can not be found.}
     \typeout{*}
     \typeout{* The fact that this package cannot be found either means that}
     \typeout{* this LaTeX release is too old or that it was installed}
     \typeout{* only in parts!}
     \typeout{*}
     \typeout{* You can use the aipproc class without this package but }
     \typeout{* you cannot use commands like \protect\includegraphics
                or \protect\resizebox}
     \typeout{* in this case.}
     \typeout{*}
     \typeout{* Please note that you will get a further error message below}
     \typeout{* about: "graphicx.sty not found" because the class will try}
     \typeout{* to load this package! Type return in response to that error.}
     \typeout{*}
     \typeout{* As a result the illustrations in aipguide will look strange.}
     \typeout{*}
     \typein{* Type <return> to continue ...}

     \gdef\resizebox##1##2{}
     \gdef\includegraphics{\textbf{graphics package missing:}}
     \problemtrue
    }

\IfFileExists{textcomp.sty}
    {
     \typeout{* ... ok }
    }
    {
     \typeout{* ... not found! }
     \typeout{*}
     \typeout{* Problem detected:}
     \typeout{*}
     \typeout{* The textcomp package, which is part of standard LaTeX}
     \typeout{* (base distribution) can not be found.}
     \typeout{*}
     \typeout{* The fact that this package cannot be found either means that}
     \typeout{* this LaTeX release is too old or that it was installed}
     \typeout{* only in parts!}
     \typeout{*}
     \typeout{* You can use the aipproc class without this package but }
     \typeout{* you will always get the error: "textcomp.sty not found"}
     \typeout{* because the class will try to load this package!}
     \typeout{* Type return in response to that error.}
     \typeout{*}
     \typein{* Type <return> to continue ...}

     \problemtrue
    }

\IfFileExists{url.sty}
    {
     \typeout{* ... ok }
    }
    {
     \typeout{* ... not found! }
     \typeout{*}
     \typeout{* Problem detected:}
     \typeout{*}
     \typeout{* The url package, which should be part of a good LaTeX}
     \typeout{* distribution, can not be found.}
     \typeout{*}
     \typeout{* Without this package you will not be able to use the \string\url}
     \typeout{* command. Try to download this package from a CTAN  host.}
     \typeout{* Refer to http://www.ctan.org and search for "url".}
     \typeout{*}
     \typein{* Type <return> to continue ...}

     \problemtrue
    }

\IfFileExists{textcase.sty}
    {
     \typeout{* ... ok }
    }
    {
     \typeout{* ... not found! }
     \typeout{*}
     \typeout{* Problem detected:}
     \typeout{*}
     \typeout{* The textcase package, which should be part of a good LaTeX}
     \typeout{* distribution, can not be found.}
     \typeout{*}
     \typeout{* Without this package you should be careful not to put math}
     \typeout{* formulas into \noexpand\section headings as these headings are}
     \typeout{* converted to UPPERCASE and might spoil your formulas.}
     \typeout{* Try to download this package from a CTAN  host.}
     \typeout{* Refer to http://www.ctan.org and search for "url".}
     \typeout{*}
     \typein{* Type <return> to continue ...}

     \problemtrue
    }

\makeatletter

\IfFileExists{natbib.sty}
    {
     \IfStandaloneCheck
       {\begingroup
        \let\@listi\relax
        \let\thebibliography\@empty
        \let\bibstyle\@empty
        \RequirePackage{natbib}
        \@ifpackagelater{natbib}{1999/05/29}
          {
           \typeout{* ... ok }
          }{
           \typeout{* ... might be too old! }
           \typeout{*}
           \typeout{* Your version of the natbib package might be too}
           \typeout{* old to be usable. This class was designed to}
           \typeout{* work with the version 7.0 dated 1999/05/28}
           \typeout{*}
           \typeout{* If problems occur download a}
           \typeout{* recent version from a CTAN host.}
           \typeout{*}
           \typeout{* Refer to http://www.ctan.org and search for "natbib".}
           \typeout{*}
           \typein{* Type <return> to continue ...}

           \global\problemtrue
          }
        \endgroup
        }{}
    }
    {
     \typeout{* ... not found! }
     \typeout{*}
     \typeout{* Serious problem detected:}
     \typeout{*}
     \typeout{* The natbib package, which should be part of a good LaTeX}
     \typeout{* distribution, can not be found.}
     \typeout{*}
     \typeout{* Without this package you will not be able to use certain}
     \typeout{* citation styles. See the aipguide documentation!}
     \typeout{*}
     \typeout{* Especially the layout for ARLO requires this package!}
     \typeout{*}
     \typeout{* Try to download this package from a CTAN  host.}
     \typeout{* Refer to http://www.ctan.org and search for "natbib".}
     \typeout{*}
     \typein{* Type <return> to continue ...}

     \problemtrue
    }

\makeatother

\ifproblem
\typein{* Type <return> to continue ...}
\else
 \ifobservation
 \else
 \fi
\fi

\makeatletter
\IfStandaloneCheck
 {
\typeout{*}
\typeout{* This document only produces terminal output.}
\typeout{*}
\stop
 }
 {
\AtBeginDocument{\relax\ifx\xfm@address@loop\@undefined
  \typeout{***************************}
  \typeout{* Oooops ... you seem to have picked up an obsolete}
  \typeout{* aipproc.cls file from a previous installation!}
  \typeout{*}
  \typeout{* Please check that LaTeX finds the right one.}
  \typeout{*}
  \typeout{* Sorry have to give up ....}
  \typeout{***************************}
  \stop
 \fi}
 }
\makeatother

\documentclass[
    ,final            
  ]
  {aipproc}

\layoutstyle{6x9}
\def\gsim{ \lower .75ex \hbox{$\sim$} \llap{\raise .27ex \hbox{$>$}} }
\def\lsim{ \lower .75ex\hbox{$\sim$} \llap{\raise .27ex \hbox{$<$}} }

\begin{document}

\title{The rate and luminosity function of Short GRBs}


\author{Tsvi Piran\footnote{talk given by Tsvi Piran}}{
  address={Racah Institute for Physics, The
Hebrew University, Jerusalem 91904, Israe}}
\author{Dafne Guetta}{
  address={ Osservatorio astronomico of Rome v. Frascati 33 00040
Monte Porzio Catone, Italy }
}

\begin{abstract}
We compare the luminosity function and rate inferred from the
BATSE short hard bursts (SHBs) peak flux distribution with the
redshift and luminosity distributions of SHBs observed by {\it
Swift}/HETE II. The {\it Swift}/HETE II SHB sample is incompatible
with SHB population that follows the star formation rate. However,
it is compatible with a distribution of delay times after the SFR.
This would be the case if SHBs are associated with binary neutron
star mergers.  The implied SHB rates that we find range from $\sim
8$ to $\sim 30h_{70}^3$Gpc$^{-3}$yr$^{-1}$. This rate is a much
higher than what was previously estimated and, when beaming is
taken into account,  it is comparable to the rate of neutron star
mergers estimated from statistics of binary pulsars. If GRBs are
produced in mergers the implied rate practically guarantees
detection by LIGO II and possibly even by LIGO I, if we are lucky.
Our analysis, which is based on {\it observed} short hard burst is
limited to bursts with luminosities above $10^{49}$erg/sec. Weaker
bursts may exist but if so they are hardly detected by BATSE or
{\it Swift} and hence their rate is very weakly constrained by
current observations. Thus the rate of mergers that lead to a
detection of a gravitational radiation signal might be even
higher.
\end{abstract}

\maketitle


\section{Introduction}

The luminosity and rate of Gamma-Ray Bursts (GRBs) is one of the
key issues in any astrophysical model. Shortly after the discovery
that GRBs are cosmological Piran \cite{piran92} and Mao and
Paczynski \cite{mp92} used the observed value of $\langle
V/V_{max} \rangle = 0.32$  and estimated, assuming that the GRB
rate is independent of redshift and that they are standard
candles, the rate of GRBs. Already at that time Piran
\cite{piran92} suggested  that if GRBs arise due to
neutron star mergers \cite{eichler89} their rate will depend on
redshift and it will follow the neutron star binary formation rate
with a logarithmic distribution of time delays, reflecting the
distribution of merger times.

In 1993 Koveliotou \cite{kouv93} noticed that the GRB distribution
can be divided into two subsets of long and short bursts with a
dividing duration of 2sec. As short bursts are also harder than
long ones \cite{dez96,kouv96}, they are denoted as Short Hard
Bursts (SHBs). Mao et al, \cite{mao94} realized that BATSE is less
sensitive to short GRBs than to long ones and pointed out that
this should be considered when trying to fit the {\it observed}
peak flux distribution to different models. Shortly afterwards
Cohen and Piran carried out the first separate analysis of the
luminosity and rates of long and short GRBs finding that the {\it
observed} SHB population is located  $z \leq 0.5$, namely they are
much nearer than the {\it observed} distribution of long ones.
Unfortunately a stubborn referee forced these authors to take out
the discussion of short GRBs from \cite{cp95} and these results
were reported only in a conference proceedings \cite{piran94} and
in Cohen's Phd thesis \cite{c96}. These finding were corroborated
shortly later by Katz and Canel \cite{kc96} and later by Tavani
\cite{ta98} who found that $\langle V/V_{max} \rangle $ of SHBs is
significantly higher than the one measured for long bursts.

With the discovery  in 1997 of GRB afterglow and the subsequent
identification of host galaxies and redshift measurements direct
estimates of the luminosity and rates were obtained for long
bursts. It was discovered that long GRBs follow the SFR and that
typical (isotropic equivalent) peak luminosities are about
$10^{51}$ergs/sec.

However, until recently no afterglow was detected from any short
burst who remained as mysterious as ever. Last spring {\it Swift}
and then HETE II detected X-ray afterglow from several short
bursts. In some cases optical and radio afterglows were detected
as well. This has lead to identification of host galaxies and to
redshift measurements. While the current sample is very small
several features emerge. First, unlike long GRBs that take place
only in star forming spiral galaxies SHBs take place also in
elliptical galaxies in which the stellar population is older. In
this SHBs behave like type Ia Supernovae. Second, the redshift and
peak (isotropic equivalent) luminosity distributions of the five
short bursts (see Table 1) confirm earlier expectations
\cite{piran94,c96,kc96,ta98,gp05} that the observed SHB population
is significantly nearer than the observed long burst population.

We \cite{gp05} (denoted hereafter GP05) have estimated, before the
launch of {\it Swift}  the luminosity function and formation rate
of SHBs from the BATSE peak flux distribution. We have shown that
the distribution is compatible with either a population of sources
that follow the SFR (like long bursts) or with a population that
lags after the SFR \cite{piran92,ando04}. If SHBs are linked to
binary neutron star mergers \cite{eichler89} the SHB rate is given
by the convolution of the star formation rate with the
distribution $P_m(\tau)$ of the merging time delays $\tau$ of the
binary system. These delays reflect the time it takes to the
system to merge due to emission of gravitational radiation.

\begin{table}[h]
  \centering
\begin{tabular}{|c|c|c|c|c|c|}
\hline
  GRB & 050509b & 050709 & 050724 & 050813 & 051221\\
\hline
  z & 0.22 & 0.16 & 0.257 & 0.7 or 1.80 & 0.5465 \\
\hline
  $L_{\gamma,iso}/10^{51}$erg/sec & 0.14 & 1.1 & 0.17 & 1.9& 3\\
\hline
\end{tabular}
\label{table1}
\end{table}
\noindent\textbf{Table 1:} The {\it Swift}/HETE II current sample
of SHBs with a known redshift.
\section{The luminosity function of the BATSE SHB sample}

As BATSE is less sensitive to short bursts than to long ones
\cite{mao94}, even an {\it intrinsic} SHB distribution that follow
the SFR gives rise to an {\it observed} distribution that is
nearer to us. Still a delayed SFR distribution (that is {\it
intrinsically} nearer) gives rise to even nearer {\it observed}
distribution \cite{gp05}. Therefore the recent {\it observed}
redshift distribution of SHBs favors the delayed model and hence
the merger scenario. Still the question was posed whether the
predicted  {\it observed} distribution is consistent with the
current sample. Gal Yam et al., \cite{GalYam05a}  suggested that
the distributions are inconsistent and hence the delayed SFR model
is ruled out. We re-examine the situation here and we show that
while a delayed distribution with  the best fit (maximal
likelihood)  parameters ia only marginally consistent
($p_{KS}=0.05-0.06$) with the current sample, a delayed
distribution with parameters within $1\sigma$ from the best fit
parameters is compatible ($p_{KS}=0.22-0.25$). We discuss the
implications of this result to the binary Neutron star merger
model and to the detection of gravitational radiation from such
mergers. We also discuss the recent suggestion of Nakar et al.
\cite{n05} that there rate of mergers producing very weak bursts
is very high.

Our data set and methodology follow \cite{gpw05,gp05}.  We
consider all the SHBs detected while the BATSE onboard trigger
\cite{pa99} was set for 5.5$\sigma$ over background in
at least two detectors in the energy range 50-300keV. These
constitute a group of 194 bursts. We assume the functional form of
the  rate of bursts (but not the amplitude). We then search for a
best fit luminosity function. Using this luminosity function we
calculate the expected distribution of {\it observed} redshifts
and we compare it with the present data. We consider the following
cosmological rates:
\begin{itemize}
    \item (i) A
rate that follows the SFR (We do not expect that this reflects the
rate of SHBs but we include this case for comparison.).
    \item (ii)  A
rate that follows the NS-NS merger rate. This rate depends on the
formation rate of NS binaries, that one can safely assume follows
the SFR, and on the distribution of merging time delays,
$P_m(\tau)$. This, in turn, depends on the distribution of initial
orbital separation $a$ between the two stars ($\tau\propto a^4$)
and on the distribution of initial eccentricities. Both are
unknown. From the coalescence time distribution of six  double
neutron star binaries \cite{champ04} it seems that
$P_m(\log(\tau))d\log(\tau)\sim$ const, implying $P_m(\tau)\propto
1/\tau$,\cite{piran92}. Therefore, our best guess scenario is a
SBH rate that follows the SFR with a logarithmic time delay
distribution.
    \item (iii)
A rate that follows the SFR with a  delay distribution
$P_m(\tau)d\tau\sim$ const.
    \item (iv) A constant rate (which is
independent of redshift.).
\end{itemize}
For the SFR  we employ  $R_{SF2}$ of Porciani \& Madau
\cite{pm01}: In models (ii) and (iii) the rate of SHBs is given
by:
\begin{equation}
\label{rate} R_{\rm SHB}(z) \propto \int_{0}^{t(z)} d\tau R_{\rm
SF2}(t-\tau) P_m(\tau) .
\end{equation}

\begin{table}[h]
\begin{tabular}{|c|c|c|c|c|c|c|}
 \hline
& Rate(z=0) & $L^*$ & $\alpha$ & $\beta$& $p_{KS}$ & $p_{KS}$ \\
   & $Gpc^{-3} yr^{-1}$ & $10^{51}$ erg/sec &   & &(z=0.7)&(z=1.8) \\
  \hline
i & $0.11^{+0.07}_{-0.04}$ & $4.6^{+2.2}_{-2.2}$
&$0.5^{+0.4}_{-0.4} $ &
$1.5^{+0.7}_{-0.5} $&$<0.01$&$<0.01$ \\
 ii &  $0.6^{+8.4}_{-0.3}$ & $2^{+2}_{-1.9}$ &
$0.6 ^{+0.4}_{-0.4}$ & $2 \pm 1$& 0.05 &0.06\\
ii$_\sigma$ & $10^{+8}_{-5}$ & 0.1 & $0.6^{+0.2}_{-0.4}$ &
$1\pm 0.5 $& 0.22 &0.25  \\
 iii & $30^{+50}_{-20}$ & $0.2^{+0.5}_{-0.195}$  &
$0.6^{+0.3}_{-0.5} $  & $1.5^{+2}_{-0.5} $&0.91 & 0.91\\
iv & $8^{+40}_{-4}$ & $0.7^{+0.8}_{-0.6}$  &
$0.6^{+0.4}_{-0.5} $  & $2^{+1}_{-0.7} $& 0.41 & 0.41\\
 \hline
\end{tabular}
\end{table}
\noindent\textbf{Table 2:} Best fit parameters Rate(z=0) , $L^*$,
$\alpha$ and $\beta$ and their $1\sigma$ confidence levels for
models (i)-(iv). Also shown are the KS probability ($p_{ks}$) that
the five bursts with a known redshift arise from this
distribution. We show two results for KS tests one with GRB
0508132 at $z= 0.7$ and the other at $z=1.8$. Case ii$_\sigma$
corresponds to case ii with an $L^*$ value lower by $1 \sigma$
than the best fit one. Other parameters have been best fitted for
this fixed number.

Following  Schmidt \cite{sch01} (see also \cite{gpw05,gp05}) we
consider a broken power law peak luminosity function  with lower
and  upper limits, $1/\Delta_1$  and $\Delta_2$, with power law
indeces, $\alpha$, $\beta$ and luminosity break $L^*$.

We use $\Delta_{1,2}=(30,100)$ \cite{gp05}. In \cite{gp06}
(denoted GP06 hereafter) we show that both limits are chosen in
such a way that a very small fraction (less than 1\%) of the {\it
observed} bursts are outside these range. Hence one cannot infer
anything from the observations on the luminosity function in this
range. Comparing the predicted distribution with the one observed
by BATSE we obtain, the best fit parameters of each model and
their standard deviation. The results are shown in table 2 and in
figs 1 and 2 of GP06.

\section{A Comparison with the current {\it Swift}-HETE II SHB sample}

We can  derive now the expected redshift distribution of the
observed bursts' population in the different models.

We assume that the minimal peak flux for detection for {\it Swift}
is $ \sim 1$ ph/cm$^{2}$/sec like BATSE (note the different
spectral windows of both detectors which makes {\it Swift}
relatively less sensitive to short  bursts).

\begin{figure}[b]
  \includegraphics[height=.3\textheight]{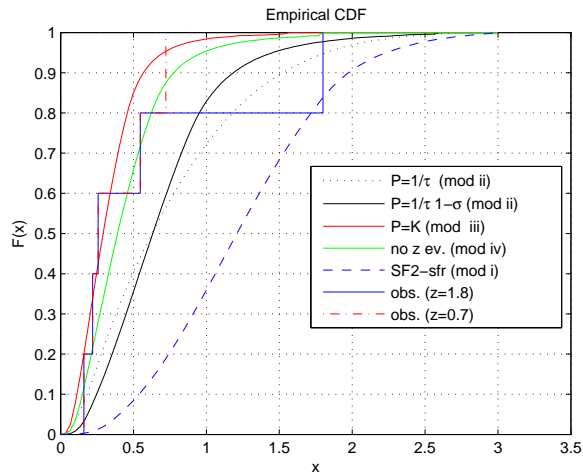}
  \caption{\label{fig5}   A comparison between
the expected integrated {\it observed} redshift distributions of
SHBs for
 models (i)-(iv) and (ii$_\sigma$) and the distribution of
known redshifts of SHBs.}
\end{figure}

Fig. \ref{fig5} depicts the expected {\it observed} integrated
redshift distributions of SHBs in the different models. As
expected, a distribution that follows the SFR, (i), is ruled out
by a KS test with the current five bursts ($p_{KS} <1$\%). This is
not surprising as other indications, such as the association of
some SHBs with elliptical galaxies suggest that SHBs are not
associated with young stellar populations.

A distribution that follows the SFR with a constant logarithmic
delay distribution, (ii), is marginally consistent with the data
($p_{KS} \sim 5-6$\%). The observed bursts are nearer (lower
redshift) than expected from this distribution. If we use the
Rowan-Robinson SFR \cite{gp05}, rather than SF2 of Porciani-Madau,
the situation is even more promising $p_{KS} \sim 10$\% (for
either z=0.7 or z-1.8). When we move to a distribution that is
1$\sigma$ away from the best fit distribution we find $p_{KS}\sim
22-25$\% and even higher for the RR SFR. Thus the suggestion
\cite{GalYam05a} that the observed sample rules out the NS merger
model (with a logarithmic merger time distribution) was somewhat
premature.  Note however, that the local rate with this model
(ii$_\sigma$) is sixteen times larger than the rate of the best
fit model, (ii). This reflects the large flexibility in modeling
the peak flux distribution.

To demonstrate the flexibility of the data we have considered two
other time delay distributions. A uniform time delay distribution
(iii) and an overall constant SHB rate (iv). Both models are
compatible with the BATSE SHB distribution and with the sample of
SHBs with a known redshift ($p_{KS}\sim $ 80\% and 40\%
respectively.). This result is not surprising. The BATSE peak flux
distribution depends on two unknown functions, the rate and the
luminosity function. There is enough freedom to chose one function
(the rate) and fit for the other.

In all  models compatible with the five bursts with a known
redshift, the {\it intrinsic} SHB rate is pushed towards lower
redshifts. The  inferred present rates, $\sim 30$, $\sim 8$ and
$\sim 10h_{70}^3$Gpc$^{-3}$yr$^{-1}$ for models (iii), (iv) and
(ii$_\sigma$) respectively, are larger by a factor ten to fifty
than those estimates earlier (assuming that SBHs follow the SFR
with a logarithmic delay with the best fit parameters GP05).  The
corresponding ``typical" luminosities, $L^*$, ranges from 0.1 to
0.7  $\times 10^{51}$erg/sec.

\section{Conclusions and Implications}

 We have repeated the analysis of fitting the BATSE SHB data to a
model of the luminosities and rates distributions. Our best fit
logarithmic distribution model is similar to the best fit
logarithmic model presented in GP05. A main new ingredient of this
work is the fact that we consider several other models. We confirm
our earlier finding that the BATSE data allows a lot of
flexibility in the combination of the rates and luminosities.

A second new ingredient of this work is the comparison of  the
best fit models to the small sample of five {\it Swift}/HETE II
SHBs. The {\it Swift}/HETE II data gives a new constraint. This
constraint favors a population of SHBs with a lower intrinsic
luminosity and hence a nearer {\it observed} redshift
distribution. It implies a significantly higher local SHB rate - a
factor of ten to fifty higher than earlier estimates. The new
observations of Swift show that the SHBs are nearer than what was
expected before and therefore, their luminosity is lower and their
local rate is higher. We stress that this new result was within
the 1$\sigma$ error of the model presented in GP05, which had a
very large range of allowed local rates and typical luminosities.

Provided that the basic model is correct and we are not mislead by
statistical (small numbers), observational (selection effects and
threshold estimates) of intrinsic (two SHB population) factors we
can proceed and compare the inferred SHB rate with the
observationally inferred rate of NS-NS mergers in our galaxy
\cite{na91}. This rate was recently reevaluated with the discovery
of PSR J0373-3039 to be rather large as $80^{+200}_{-66}$/Myr.
Although the estimate contains a fair amount of uncertainty
\cite{ka04}. If we assume that this rate is typical and that the
number density of galaxies is $\sim 10^{-2}$/Mpc$^{3}$, we find a
merger rate of $800^{+2000}_{-660}$/Gpc$^{3}$/yr. Using a beaming
factor of 30-50 for short bursts \cite{be05} this rate implies a
total merger rate of $\sim 240-1500$/Gpc$^{3}$/yr for the three
cases (iii), (iv) and (ii$_\sigma$). The agreement between the
completely different estimates is surprising and could be
completely coincidental as both estimates are based on very few
events.

If correct these estimates are excellent news  for gravitational
radiation searches, for which neutron star mergers are prime
targets. They imply that the recently updates high merger rate,
that depends mostly on one object, PSR J0737+3039, is valid. These
estimate implies one merger event within $\sim 70$Mpc per year and
one merger accompanied with a SHB within $\sim 230$Mpc. These
ranges are almost within the capability of LIGO I and certainly
within the capability of LIGO II. If correct these estimates of
the rate are excellent news  for gravitational radiation searches,
for which neutron star mergers are prime targets.

To conclude we stress that we have assumed that the luminosity
function has a lower limit of $L^*/30$. This was just because even
if such a limit does not exists weaker bursts would be barely
detected. The current peak flux distribution of BATSE burst cannot
confirm (or rule out) the existence of such population (note
however, that Tanvir et al., \cite{tavnir05} suggest that such a
population exists on the basis of the angular distribution of
BATSE SHBs). If such weak bursts exist then, of course, the
overall merger rate will be much larger \cite{n05}. Such events
will provide such a high rate that soon LIGO I will begin to
constrain this possibility.




\bibliographystyle{aipproc}   

\end{document}